\documentclass[showpacs,amsmath,amssymb,amsfonts]{revtex4}

\usepackage{enumerate}
\usepackage{graphicx}
\usepackage{dcolumn}
\usepackage[usenames]{color}
\usepackage{ulem} 

\begin{document}
\title{Comment on ``Third Law of thermodynamics as a key test of generalized entropies''}

\author{G. Baris Bagci}
\email{gbb0002@hotmail.com} \affiliation{Department of Materials Science and Nanotechnology Engineering, TOBB University of Economics and Technology, 06560 Ankara, Turkey}
\author{Thomas Oikonomou}
\affiliation{Crete Center for Quantum Complexity $\&$ Nanotechnology, 
      Department of Physics, University of Crete, 71003 Heraklion, Hellas}

\affiliation{Department of Physics, Faculty of Science, Ege University, 35100 Izmir, Turkey}


\begin{abstract}

Bento \textit{et al.} [Phys. Rev. E 91, 022105 (2015)] state that the Tsallis entropy violates the third law of thermodynamics for $q \leq 0$ and $0<q<1$. We show that their results are valid only for $q \geq 1$, since there is no distribution maximizing the Tsallis entropy for the intervals $q \leq 0$ and $0<q<1$ compatible with  the system energy expression.  
\end{abstract}

\pacs{05.70.-a}
\newpage \setcounter{page}{1}

\maketitle


In Ref. \cite{Bento}, the Tsallis entropy i.e., $S = \frac{\sum_{\lambda} p_{\lambda}^{q}-1}{\left( 1-q\right) }$, is considered and concluded that it violates the third law of thermodynamics for $q<1$. In particular, to this aim, Ref. \citep{Bento} makes use of the Tsallis entropy constrained by the normalization $\sum_{\lambda} p_{\lambda}=1$ and the system energy definition given by $U = \frac{\sum_{\lambda} p_{\lambda}^{q} E_{\lambda}}{\sum_{\lambda} p_{\lambda}^{q}}$ as can be seen from Eqs. (2) and (7) in Ref. \cite{Bento}. $\lambda$ denotes the number of accessible microstates beginning with the state zero while $n$ denotes the same quantity starting from one. However, the equilibrium probability distributions are implicitly considered, and not explicitly analysed in Ref. \cite{Bento}. For an explicit analysis, one notes that the corresponding Tsallis equilibrium distribution, through entropy maximization procedure $\frac{\partial S}{\partial p_{\lambda}}=0$ with the aforementioned constraints, reads \cite{Tsallis}

\begin{equation}\label{distr}
p_{\lambda} =  \left[  1+(q-1)\frac{\beta_{\lambda}}{\sum_{\kappa} p_{\kappa}^{q}}\left(E_{\lambda}-U \right)        \right]_{+}^{\frac{1}{1-q}}
\end{equation}

\noindent apart from the partition function where $\left[ a\right]_{+}= \text{max} \left\lbrace 0,a\right\rbrace $ to ensure the non-negativity of the probabilities. Note that the probabilities $p_{\lambda}$ are zero whenever the condition $\frac{\beta_{\lambda}}{\sum_{\kappa} p_{\kappa}^{q}}\left( E_{\lambda}-U\right) \leq 1/(1-q)$ is satisfied for the interval $q\in(0,1]$ due to the non-negativity requirement.

For the above equilibrium distribution to maximize the Tsallis entropy, the following condition should also hold \cite{note} 

\begin{equation}\label{criterion}
\frac{\partial^{2} S}{\partial p_{\lambda}^{2}}=-q p_{\lambda}^{q-2} < 0,
\end{equation}

\noindent which at once shows that the distribution given in Eq. (1) does not maximize the Tsallis entropy for $q \leq 0$. Therefore, it cannot be used as an equilibrium distribution for $q \leq 0$. In Ref. \cite{Bento}, the authors state that the third law is violated for $q \leq 0$ in the case of the Tsallis entropy although there is no known equilibrium distribution for the aforementioned values of the Tsallis parameter $q$. Moreover, the substitution of the equilibrium distribution $p_{n}$ (note that $n =1,2,3...$ as in Ref. \cite{Bento}) in Eq. (1) into Eq. (2) shows that the equilibrium distribution can be considered only for $1 \leq q < 2$ if one imposes the limit $\beta_{n}\rightarrow+\infty$ for any $n$ starting from one, since the limit necessary for the third law i.e. $\beta_{n}\rightarrow+\infty$ and Eq. (2) are both satisfied only for $q=1$ in the interval $q\in(0,1]$ due to the cut-off condition $\frac{\beta_{n}}{\sum_{\kappa} p_{\kappa}^{q}}\left( E_{n}-U\right) \leq 1/(1-q)$ where $E_{0}=U$ so that $\left( E_{n}-E_{0}\right) >0$ for the third law as shown in Ref. \cite{Bento}. With the chosen constraints in Ref. \cite{Bento}, one can check any equilibrium feature of the Tsallis entropy only for $1 \leq q < 2$. Therefore, the results in Ref. \cite{Bento} are not valid for the interval $0 < q < 1$.

Another issue stemming from the treatment in Ref. \cite{Bento} considering the inclusion of the interval $q\in(0,1]$ is that the iff-condition in Eq. (3)  in Ref. \cite{Bento} concerning the 3rd law of thermodynamics is violated due to the cut-off of the equilibrium distribution in Eq. (1), since   $p_{n}$ may become zero both in the limit $\beta_{n}\rightarrow+\infty$ and when $\frac{\beta_{n}}{\sum_{\kappa} p_{\kappa}^{q}}\left( E_{n}-U\right) \leq 1/(1-q)$ as well. As a result, one can choose $p_0$ as one and all the other $p_n$'s as zero thereby making the whole entropy equal to zero without guaranteeing the condition $\beta_{n}\rightarrow+\infty$ in a unique manner as required by the third law of thermodynamics.

To sum up, the calculations in Ref. \cite{Bento} for the third law presuppose certain limits to be taken in terms of the equilibrium distributions just as is the case with ordinary Boltzmann-Gibbs entropy. However, through the precedence, one should first have the appropriate equilibrium distributions maximizing the Tsallis entropy for the related intervals. The Tsallis entropy together with the system energy expression used in Ref. \cite{Bento} allows the Tsallis distributions only for $1 \leq q < 2$. Finally, we note that the steady state Tsallis distributions arising from non-equilibrium frameworks are beyond the scope of this comment, since they do not have to obey the third law of thermodynamics which is an equilibrium feature.

\section*{Acknowledgements}

We thank G. M. Viswanathan, M. G. E. da Luz and R. Silva for fruitful correspondence. This work was partially supported by the European Union's Seventh Framework Programme (FP7-REGPOT-2012-2013-1) under grant agreement n$^o$ 316165.


\begin{thebibliography}{00}
\bibitem{note} The same condition i.e. $\frac{\partial^{2} S}{\partial p_{\lambda}^{2}}< 0$  yields $- p_{\lambda}^{-1} < 0$ for the Boltzmann-Gibbs entropy given by $S = -\sum_{\lambda} p_{\lambda}\ln p_{\lambda}$. For the third law of thermodynamics as an equilibrium feature, the criterion $- p_{\lambda}^{-1} < 0$ should be satisfied for the canonical distribution $p_{\lambda} = e^{- \beta E_{\lambda}}$ even in the $\beta \rightarrow+\infty$ limit. Using the canonical distribution in this particular limit assuming positive microstate energies, one obtains $- \infty <0$ which shows that the criterion $\frac{\partial^{2} S}{\partial p^{2}}< 0$ is satisfied even when $\beta$ goes to $+\infty$.
\end{thebibliography}
\end{document}